# To What Extent Would E-mall Enable SMEs to Adopt E-Commerce?

Adel A. Bahaddad[1,2], Rayed AlGhamdi[2] & Luke Houghton[3]

[1] Information and Communication School, Griffith University, Brisbane, Australia

[2] Faculty of Computing & IT, King AbdulAziz University, Jeddah, Saudi Arabia

[3] Business School, Griffith University, Brisbane, Australia

Correspondence: Adel A. Bahaddad, Institute of Integrated and Intelligent Systems, Griffith University, Brisbane, Room 1.45-N34, 170 Kessels Road, Nathan Qld 4111, Australia. Tel: 61-737-353-765. E-mail: Adel.Bahaddad@Grifiithuni.edu.au

## Abstract

This paper presents findings from a study of e-commerce adoption by Small and Medium Enterprises (SMEs) in Saudi Arabia. Only tiny number of Saudi commercial organizations, mostly medium and large companies from the manufacturing sector, in involved in e-commerce implementation. The latest report released in 2010 by The Communications and Information Technology Commission (CITC) in Saudi Arabia shows that only 8% of businesses sell online. Accordingly new research has been conducted to explore to what extent electronic mall (e-mall) would enable SMEs in Saudi Arabia to adopt and use online channels for their sales. A quantitative analysis of responses obtained from a survey of 108 SMEs in Saudi Arabia was conducted. The main results of the current analysis demonstrate the significant of organizational factors, and technology and environmental factors. Interestingly, traditional & cultural factors have no significance in this regard.

## 1. Introduction

Many countries rely primarily on SMEs to support the economy because they represent a huge number of activities and attract a significant number of workers that assist in developing national and global economies.

The number of micro-enterprises and SMEs in the European Union (EU) is estimated to be 19.3 million, and this represents about 90% of EU enterprises. In addition, they provide around 65 million jobs, representing two-thirds of all employment (The Commission of the European Communities 2003). In Japan, 81% of all employment comes from SMEs, and in Okinawa, over 95% of all employment comes from such enterprises (Samiha & Sanjay, 2002). In Saudi Arabia, the SMEs represent around 95% of all Saudi enterprises (Al-Mahdi, 2009). Thus, SMEs are key players that are rapidly influencing economies in many countries around the world. This will lead people to rethink the manner in which the performance of SMEs can be elevated at the local, regional, and global levels.

Saudi Arabia is one of the developed countries in terms of electronic trading and the country's growth in Internet uptake. It is thirty-eighth globally, and fourth in the Arab world, in terms of network readiness, environment and infrastructure. Therefore, it is important to consider the obstacles to activating the e-Mall framework in a country that contains all of these facilities. The SMEssector provides at least 60% of jobs in the private sector (Bundagji, 2005). Many of these companies cannot offer their products online, mainly because of the cost of construction, maintenance and management of a website on the Internet. This leads us to think about what tools and applications could be used to support this sector electronically and enable SMEs to provide their services over the Internet.

## 2. Literature Review

### 2.1 Small and Medium Enterprises SMEs

Generally, SMEs are defined as non-subsidiary and sovereign firms in which the number of employees is less than a stated number (this number varies across nations). For example, in the USA, SMEs are defined as firms with less than 500 employees; in the European Union, the SME limit is set to 250 workers; while in other nations, the SME limit is set at 200 employees. Small firms are supposed to have fewer than 50 employees, while micro-enterprises have ten employees at most (Sara & Dylan, 2006). SMEs can also be defined by their financial assets. Moreover, SMEs can be described according to whether they are independently owned and operated, as in the USA. The American definition assumes that SMEs are independently owned and operated and that they are not dominating their field in the overall marketplace (Bose & Sugumaran, 2006). SMEs in Saudi Arabia are categorised as small (with less than 60 employees) and medium sized between 60 and 99 employees; according to the Council of Saudi Chambers (CSC 2010).

Several studies have concentrated on the unique characteristics of SMEs that make them different from large companies. Most of these studies have focused on the procedures and systems that have been successfully adopted in the large enterprises and the fact that they may not produce similar results when adopted by SMEs (Bose & Sugumaran, 2006). Therefore, some SME characteristics should be evident, especially when compared with the large enterprises. Firstly, SMEs do not need to contain sufficient technical experience and human resources to develop complex solutions and to assume technical improvements (Barry & Milner, 2002). Secondly, SMEs contain small groups of people, and they are influenced by the opinions of the business owner and his/her background (Bose & Sugumaran, 2006). Furthermore, SMEs do not have centralized structures with CEOs correctlymaking the critical decisions (Mintzberg, 1979). Moreover, many SMEs avoid complex programs and applications. In addition, they have suffered from poor organization (Tetteh & Burn, 2001), so they prefer to employ workers with general labour skills rather than specialists because of difficulties in keeping and attracting skilled staff (Blili 1993; Gable 1991; Nooteboom 1988). Finally, SMEs are unable to absorb the results of failures in technology adoption; therefore, they are wary of new technology (Thong et al. 1994). Some business characteristics are influenced by the application and use of the IT perspective in SMEs. These characteristics are size, age sector, focus of commerce, and level of IT knowledge between others (Auger & Gallaugher, 1997).

Adoption of the online trading framework will also boost the global economy. Dinesh and Jaideep (2001) suggested that the following factors contribute to increasing the use of online trading in SMEs. These factors are CEO's awareness of the relative benefits expected from online trading; range of compatibility of the online trading perspective with the company's plan; administrative time that is required to plan and implement the company decision; level of necessity of the company regarding the online trading perspective; competition environment of the company; setup cost; CEO's enthusiasm toward this project.

## 2.2 Electronic Mall (E-Mall)

E-Mall is a platform for physical and virtual organizations using a uniform system for purchases, supplies, and deliveries (Asfoura*et al.*, 2009). E-Malls provide a digital environment that facilitates the meeting of buyers and sellers in the same place (Timmers, 1998). While Zimmermann (1997) acknowledges that an E-Mall is a regional market with multi-national products from the same region, an e-Mall may also contain products from multiple contiguous regions. It delivers a platform for e-commerce and all other forms of information exchange and communications among commercial and private participants (Laudon & Traver 2010). It is a direct broker facilitating deals among customers and companies, and supervises the transactions that take place (Asfoura*et al.*, 2009). It can be said that e-Malls are giant frameworks that present product categories electronically from multiple sellers.

E-Malls or electronic markets are based on the transaction fee revenue model. The revenue model involves five models: advertising, subscription, sales, affiliates, and transaction fees (Laudon & Traver 2010). Transaction fees can be defined as fees a company (such as eBay.com) receives for enabling or executing a transaction (Laudon & Traver 2010). This website receives the fee for each transaction done by a customer or vendor if a vendor is successful in selling the items.

Globally, e-Malls are one of the most important purchase sites. Its diffusion was due to support by SMEs, which could not invest through the Internet alone and were unable to reach a large segment without entering into a framework such as this. In addition, e-Malls are useful, in that they increase chances of accessing national and worldwide markets.

The last decade has been a good time for the Internet. Online shopping has increased speedily. Netscape presented SSL encryption of online data transfers, which has come to be fundamental for secure online shopping websites. Then, the websites that were established at that time can be termed the nucleus of commercial sites on the Internet (such as eBay, which was launched in 1996, and Amazon, which appeared in 1995). Amazon and eBay were the early players in the e-Mall frameworks field. They have been met with many huge successes by inviting SMEs to work under their frameworks to later increase their website reputation and revenue rapidly.

*2.3 The Saudi Arabian Market*

The average of annual population growth rate in the Kingdom of Saudi Arabia has reached3% over the past 10 years. This seems to be playing a role in making Saudi Arabia the most dynamic retail markets and most notably in the Middle East (AME info 2008; Alghamdi, Drew & Al-Gaith 2011). "With an estimated population of 24.9 million and a per capita GDP of US$24,581 in 2008, Saudi Arabia is the largest and one of the richest retail markets in the Middle East" (ACG, 2009).

According to the latest report issued by Saudi Alhokair group, the wholesale and retail trade grew at a compound annual growth rate of 5.8% in the past 10 years. In 2010 the retail trade volume exceeded SAR90 billion (US$1= KSR 3.75), although it was only expected to reach up to 70 billion. It is expected that the volume of retail trade to SAR 130 billion in 2012(AME info, 2008). In 2011, the size of the retail market in Saudi Arabia estimated greater than KSR 160 billion dominated by small and medium size companies accounting for more than85% of market share (Habtoor, 2011).

On the ICT market side, the number of Internet users in Saudi Arabia grew from around 1 million to an estimated 12.5 million in the period between 2001 and the end of the first half of 2011. This number represents 44% of the population (MCIT 2011). The growth in broadband availability, decrease costs of personal computers laptops, increase public awareness and IT literacy, availability of local content on the Internet, increase of e-services such as online banking and e-government applications are all contribute to the notable growth of Internet users in Saudi Arabia (MCIT 2011).

Although the country has the largest and fastest growing ICT marketplace in the Arab region, e-commerce activities have not progressed at a similar speed (Alghamdi, Drew & Alkahlaf 2011; AlGhamdi, Drew & Alshehri 2011). CITC's IT Report 2010 reiterated that e-commerce in Saudi Arabia is still in its early stages. In particular, most Saudi retail chains have yet to establish an online channel, and only8% of Saudi businesses sell online (CITC 2010; AlGhamdi, Drew & Alhussain 2012).

The spending on online shopping in Saudi Arabia is growing. The online retail sector size is estimated to about SAR 3 billion (US$1= SAR 3.75). This figure represents about 20% of the total electronic trading in Saudi Arabia. The average value of what a consumer pays for each online purchase is about SAR400 (Hamid 2011). This growth in spending of the community of online customers combined with the slowness of retailers in introducing an online sales channel in Saudi Arabia is an indicator that retailers in Saudi Arabia are not realizing the importance of online retail yet. However, encouragingly "roughly half of businesses in Saudi Arabia have plans to adopt e-commerce in the soon future. In terms of their target customers' revenue share, B2B was estimated at 54%, B2C at 28%, and B2G at 18%" (CITC 2010).

*2.4 Factors Influencing Electronic Market Diffusion*

The literature identified various factors influencing the adoption and diffusion of e-commerce by SMEs. These factors can be categorised into three groups: people and organisational factors, technology and environmental factors, and traditional and cultural factors. See Table 1 below.

Table 1. Summary of the factors influencing SMEs to adopt e-commerce as identified in the literature

| | Factor | Source |
|---|---|---|
| People and organisational factors | Lack of interest and awareness | (Koh & Maguire, 2004; Scupola, 2002; Tan, Macaulay, & Scheurer, 2006; Taylor & Murphy, 2004; Bose & Sugumaran, 2006) |
| | Educational background | (MacGregor, 2004; Xu & Quaddus, 2004) |
| | Previous IT experience | (Eastin, 2002; MacGregor, 2004) |
| | Lack of resources and IT skills | (Taylor & Murphy, 2004; Dixon *et al.*, 2002; Fillis*et al.*, 2004) |
| | Firm size | (Mirchandani & Motwani, 2001; Bose & Sugumaran, 2006; MacGregor, 2004) |
| | Business sector | MacGregor, 2004 |
| | Electronic facilitator | (Martin & Matlay, 2003) |
| | Online trading advantage | (Bose & Sugumaran, 2006; AlGhamdi*et al.* 2011) |

| | Applying an innovation | (Mustonen-Ollila & Lyytinen 2003; Vézina*et al.* 2003; Koh & Maguire 2004) |
|---|---|---|
| Technology and environmental factors | Telecommunication infrastructure | (Tan, Macaulay & Scheurer 2006) |
| | Website security | (Bose & Sugumaran 2006) |
| | Delivery system | (Alfuraih 2008) |
| | Safety payment method | (Deck 1997) |
| | E-Mall's website characteristics | (Geng*et al.* 2003; Bose & Sugumaran 2006) |
| Traditional & cultural | Arabic social structure | (Aljefri, 2003) |
| | The level of development between large and small cities | (Aljefri, 2003; AlGhamdi*et al.*2011) |
| | Constraints that affect the application of e-Mall framework technology | (Aljefri, 2003; Nahlah, 2001) |

***People and organisational factors***- These factors include lack of interest and awareness, educational background, Previous IT experience, Lack of resources and IT skills, Firm size, Business sector, and Online trading advantage. Several studies mentioned that when SMEs lack of awareness about how new technologies could enhance their businesses influence negatively the adoption (Koh & Maguire 2004; Scupola 2002; Tan, Macaulay & Scheurer 2006; Taylor & Murphy 2004; Bose & Sugumaran 2006). A decision to adopt a new technology in SMEs is also affected by the executive chairman's level of education (MacGregor 2004; Xu & Quaddus 2004). Similarly with previous IT experience, SMEs who have previous experience with IT technology are more likely to adopt e-commerce (Eastin 2002; MacGregor 2004). On the other hand, a lack of resources and information technology skills is one of the key factors that hinder SMEs to adopt e-commerce (Taylor & Murphy, 2004; Dixon et al., 2002; Fillis & Wagner 2004). In addition, a company's size represents another influence on the decision to adopt e-commerce (Fariselliet al., 1999; Mirchandani & Motwani, 2001). The type of business or business sector also plays a role in determining the adoption of e-commerce. MacGregor (2004) indicated that service-oriented companies are more likely to adopte-commerce than manufacturing firms. This is because, to somehow, the former one does not involve with physical shipments. The relative of advantages is an encouraging factor for businesses to adopt e-commerce. When businesses consider e-commerce as tool to expand their marketplace, reduce costs and so forth, they are more likely to adopt it (Bose & Sugumaran 2006; AlGhamdiet al. 2012).

Technology and environmental factors- These factors include applying an innovation, telecommunication infrastructure, website security, delivery system, safety payment method, and e-Mall's website characteristics. Innovation factors such as clarity of vision and the ability to experiment have significant conjunction with technology adoption (Mustonen-Ollila & Lyytinen 2003; Roger 1995). The e-Mall framework is a model for collaborative e-Mall offering a higher degree of innovation than other e-commerce business models (Vézinaet al., 2003). Therefore, SMEs are less confident to adopt technology that has not proven successful elsewhere, but they are likely to apply the software or applications that have already been tested comprehensively by other organisations, or in other countries (Koh & Maguire 2004). In addition, telecommunication infrastructure not only a factor influencing a decision but is a requirement to enable e-commerce (Tan, Macaulay & Scheurer 2006). Beside the ICT infrastructure as requirement for e-commerce, delivery systems (Alfuraih 2008) and online payment methods (Deck 1997) play significant role in adoption. Security is also a major concern for businesses. One of the key aspects of Web services management is to ensure that services can be provided and accessed according to the clear security policies and does not constitute a significant danger to the organisations during implementation (Bose & Sugumaran 2006).

***Traditional and cultural factors***- A successful online trading applications in a particular region may not necessarily mean that they are successful in another region. Cultures and traditions play a significant role in this regard (Curbera et al., 2003). The level of development between the larger, metropolitan cities and rural areas is radically different. These significant variations could point to the positive side of e-market growth, and enhance consumers' sophistication in areas where there are fewer commercial Malls (Aljefri, 2003).

E-commerce in Saudi Arabia is associated with its global counterpart, and with the country's level of economic development, technical, and social traditions. These paradoxical factors play a significant part in encouraging the promotion of e-commerce (Aljefri, 2003). In Saudi Arabia, two restrictions are the social aspect of visiting shopping Malls, and the importance of displaying and touching goods before the procurement process. Many Saudi families are interested in going Mall centres for enjoyment, window shop, and enhance the social side of family life (Aljefri 2003). Furthermore, most people are still not familiar with it. Thus, many people still not familiar with buying online and the old buying habit dying hard (AlGhamdiet al. 2011). Consumers mostly have concerns with the lack of physically inspecting a product in one's hand in online purchases (AlGhamdiet al. 2011). Also, they have concerns with a lack of in-person communication with sellers. As a result, it has been asserted that this would be one of the arguments against diffusing the e-Mall (Qasrawi, 2001). Moreover, Aljefri (2003) argued that there are other psychological obstacles because the modern economy depends on the exchange of money, which means that consumers are familiar with exchanging cash funds, and in face-to-face interactions.

## 3. Methodology

In this paper, a literature review was conducted to come up with a list of issues influencing SMEs to adopt e-commerce. A questionnaire survey is used to gain statistical information about the significance of these issues among SMEs in Saudi Arabia to adopt e-mall solution. The survey questions are designed in English with an Arabic translation version being available, so that the participant can opt for the most familiar language. Around 140 paper copies of the questionnaire forms were distributed in person to SMEs in Jeddah (the main economic city in Saudi Arabia) during April to June of 2011. Potential participants were selected via the "snowballing" approach where some participants were initially approached, and then they would be asked to recommend others who might be willing to participate, and so on. A total of 94 completed forms were returned, giving a response rate of around 67%.

Electronic copies of the questionnaire forms were also kept on the Survey Monkey website. The authors collected the email addresses of 400 SMEs that were members of the Jeddah chambers of commerce. Invitations to participate online were sent via email to these 400 addresses, but around 100 were returned because the addresses were invalid. In total there were 300 businesses that potentially could choose to participate in the online survey via the Survey Monkey website. At the time of writing (Sep 2011), 14 had done so, implying a response rate of 4.6%.

The number of business companies that represent the corporate sector in Saudi Arabia is 848,500 commercial enterprises by the end of 2010 (AL-Riyadh, 2010). Also, according to SAMA's report that represent the SME represent 90% (SUSRIS, 2011), the estimate number could be up to 763,650 within the SME sector. Based on this information, the calculation of the sample size can be represented this sector, according to the following equation, and the symbols interpret as follows:

$$X = Z(C/100)^2 *r(100-r)$$
$$n = N*x / ((N-1)*E2 + x)$$
$$E = \text{Sqrt} [(N - n)x/n(N-1)]$$

Figure 1. Sample size equation

Source: (Raosoft, 2004)

- (N) Represents the companies' number which is 763,650.
- (Z(c/100)) Is the critical value for the confidence level c.
- (r) Is the fraction of responses that you are interested in.
- (c) Is the level of confidence, which was represented 95% in this study.
- (n) Is the sample size output
- (E) Represents the margin of error, which was calculated as 10 %.

The number of sample size output was 97 responses, within an error margin of 10% and 95% level of confidence. The samples have been collected in this study was 110 responses, and this was appropriate number for this study based on the sample size equation result.

Table 2. Attributes of SMEs participants

| Category of SMEs | No. | % |
|---|---|---|
| All participating | 110 | 100 |
| *Company size* | | |
| Less than 10 | 14 | 12.7% |
| 10–59 | 36 | 32.7% |
| 60–100 | 60 | 54.5% |
| *Product type* | | |
| Intangible products | 6 | 5.5% |
| Tangible products | 80 | 72.7% |
| Both | 24 | 21.8% |
| *Business type* | | |
| Real Estate and Public Services | 6 | 5.45% |
| Books/Learning Resources | 6 | 5.45% |
| Electronics and Electrical | 18 | 16.4% |
| Car parts and Accessories | 10 | 9.1% |
| Beauty Products | 10 | 9.1% |
| Clothing | 14 | 12.7% |
| Furniture | 10 | 9.1% |
| Jewellery | 8 | 7.28% |
| Medicines and Medical Supplies | 2 | 1.8% |
| Household Items, Tools, and Toys | 8 | 7.27% |
| Tickets, Hotel, Car Rental | 6 | 5.45% |
| Computer related | 6 | 5.45% |
| Grocery | 6 | 5.45% |
| *Internet Access Availability* | | |
| Yes | 104 | 94.5% |
| No | 6 | 5.5% |
| *Website* | | |
| Business already has a website | 74 | 67.2% |
| Company plans to have a website | 8 | 7.3% |
| No website | 28 | 25.5% |
| *Purpose of the website* | | |
| No identified purpose | 30 | 27.3% |
| Presenting product information | 60 | 54.5% |
| Customer service and support | 52 | 47.3% |
| Distribute offers and new products | 36 | 32.7% |
| Sale of part, or all, of company products | 26 | 23.6% |

## 4. Results and Data Analysis

This section presents a summary and analysis of the responses collected to date from 110 participants. Enterprises with less than 10 employees represent 12.7% of the sample, compared with 32.7% for enterprises with 10-59 employees and 54.5% for Medium enterprises (60-100 employees). About 95% of the businesses in the sample have access to Internet. More than two-third of the businesses (67.2%) already own organizations' websites. The purposes of an organization website is various from a company to another. Only 23.6% of the businesses, which own website, conduct part or full online sales for their products. While the others making the website for other purposes; 54.5% to present information about products, 47.3% for customer services and support purposes, 32.7% to distribute services and support, while 27.3% with no identified purposes. Table 2 below presents full attributes of the research participants.

Table 3 present details of the analysis for all factors for SMEs participants. Significant factors were recognized using standard T-tests. Only 9 factors that were considered (Table 3) and significant at the 0.05 level based upon the T-tests. These factors include comes from the two first groups: people & organizational factors and technology & environmental factors. Interestingly, none of the factors in the third group, traditional & cultural factors, were identified significant.

Table 3. Data analysis from all respondents showing factors listed in order of descending mean value for each group

| | | N | Mean | Std. Deviation | t-value | Level of Sig. (2-tailed) | 95% Confidence Interval of the Difference | |
|---|---|---|---|---|---|---|---|---|
| | | | | | | | Lower | Upper |
| People & organisational factors | Online trading advantage | 108 | 2.3333 | 0.98588 | 24.596 | .000 | 2.1453 | 2.5214 |
| | Firm size | 108 | 2.2870 | 1.10268 | 21.554 | .000 | 2.0767 | 2.4974 |
| | Business sector | 108 | 2.0000 | 0.84278 | 24.662 | .000 | 1.8392 | 2.1608 |
| | Previous IT experience* | 108 | 1.8889 | 0.92052 | 21.325 | .000 | 1.7133 | 2.0645 |
| | Lack of resources and IT skills* | 108 | 1.7963 | 0.78251 | 23.856 | .000 | 1.6470 | 1.9456 |
| | Lack of interest and awareness* | 108 | 1.7037 | 0.65936 | 26.853 | .000 | 1.5779 | 1.8295 |
| | Educational background* | 108 | 1.3333 | 0.61142 | 22.663 | .000 | 1.2167 | 1.4500 |
| Technology & environmental factors | E-Mall's website characteristics* | 108 | 1.7037 | 0.89921 | 19.690 | .000 | 1.5322 | 1.8752 |
| | Applying an innovation | 108 | 1.6852 | 0.74443 | 23.525 | .000 | 1.5432 | 1.8272 |
| | Telecommunication infrastructure | 108 | 1.5556 | 0.87897 | 18.392 | .000 | 1.3879 | 1.7232 |
| | Delivery system* | 107 | 1.5327 | 0.66329 | 23.903 | .000 | 1.4056 | 1.6598 |
| | Website security* | 106 | 1.4528 | 0.77006 | 19.424 | .000 | 1.3045 | 1.6011 |
| | Safe payment method* | 108 | 1.3333 | 0.64126 | 21.608 | .000 | 1.2110 | 1.4557 |
| Traditional & cultural | Arabic social structure | 108 | 2.4630 | 1.01784 | 25.147 | .000 | 2.2688 | 2.6571 |
| | Constraints that affect the application of e-Mall framework technology | 107 | 2.3242 | 0.98558 | 22.576 | .000 | 2.1247 | 2.5011 |
| | The level of development between large and small cities | 107 | 2.1028 | 1.13216 | 19.212 | .000 | 1.8858 | 2.3198 |

* significant at the 0.05 level for this group of all respondents

Among the people and organizational factors, good previous IT experiences in the organization and educational background of the management play positive role in adopting e-mall. In contrast, lack of IT resources and skills in an organization, and lack of interest and awareness among managers/owners negatively influence the decision to adopt e-mall solution.

Significant factors among the technological and environmental category include e-mall website characteristics and its security, safe payment methods, and delivery system. An e-mall website that enhanced with effective tools of marketing and easy to use plays positive role in adopting e-mall solution. However, the level of security in this website is concerning the businesses. This is, to some extent, related to safe payment methods available. The more available trustworthy and secure online payment options play highly and positively influential role.

The level of telecommunication and logistics infrastructure in the country also play significant role in the decision to adopt e-mall solution. Doing business online required a minimum level of ICT and logistics infrastructure to enhance businesses to run their businesses. High level of an organization e-readiness would not help to do business online without the support of ICT and delivery infrastructures in the country.

## 5. Conclusion

This paper reviewed the literature and come up with 17 factors influencing SMEs to adopt e-commerce. These factors were categorised into three groups: people and organizational, technological and environmental, and traditional and cultural. A questionnaire survey was used to gain statistical information about the significance of these issues among SMEs in Saudi Arabia to adopt e-mall solution. The data collected from 110 SMEs in Saudi Arabia. Only 9 factors were identified significant. These factors are previous IT experiences; educational background; lack of resources and IT skills; lack of interest and awareness; e-Mall's website characteristics and its security; trustworthy and secure payment methods; telecommunication and logistics infrastructures.

## References

ACG (Alpen Capital Group). (2009). *GCC Retail Industry*. Alpen Capital Group.

Alfuraih, S. (2008). *E-commerce and E-commerce Fraud in Saudi Arabia: A Case Study*. In 2nd International Conference on Information Security and Assurance (pp. 176-80). Busan, Korea.

AlGhamdi R., Nguyen A., Nguyen J., & Drew S. (2012). Factors Influencing E-Commerce Adoption by Retailers in Saudi Arabia: A Quantitative Analysis. *International Journal of Electronic Commerce Studies (IJECS), 3*(1), 85-100.

AlGhamdi, R., Drew, S., & Al-Ghaith, W. (2011). Factors Influencing Retailers in Saudi Arabia to Adoption of Online Sales Systems: a qualitative analysis. *Electronic Journal of Information System in Developing Countries (EJISDC), 47*(7), 1-23.

AlGhamdi, R., Drew, S., & Alhussain, T. (2012). A Conceptual Framework for the Promotion of Trusted Online Retailing Environment in Saudi Arabia. *International Journal of Business and Management, 7*(5), 140-149. http://dx.doi.org/10.5539/ijbm.v7n5p140

AlGhamdi, R., Drew, S., & Alkhalaf, S. (2011). *Government Initiatives: The Missing Key for E-commerce Growth in KSA*. In International Conference on e-Commerce, e-Business and e-Service (vol. 77, pp. 772-775). Paris, France.

AlGhamdi, R., Drew, S., & Alshehri, M. (2011). Strategic Government Initiatives to Promote Diffusion of Online Retailing in Saudi Arabia'. In P Pichappan (ed.), *Sixth International Conference on Digital Information Management* (pp. 217-222). Melbourne, Australia. http://dx.doi.org/10.1109/ICDIM.2011.6093333

AlGhamdi, R., Nguyen, A., Nguyen, J., & Drew, S. (2011). Factors Influencing Saudi Customers' Decisions to Purchase from Online Retailers in Saudi Arabia: A Quantitative Analysis. *IADIS International Conference e-Commerce 2011*(pp. 153-161). Roma, Italy.

Aljefri A. M. (2003). *Impact of electronic commerce on the community in the Kingdom of Saudi Arabia*. Retrieved from www.minshawi.com/other/e-commerce-in-sa.htm

Al-Mahdi, H. (2009). *Supporting SME's by Universities: An Empirical Study in Saudi Arabia towards Building a Conceptual Model for Best Practices*. Retrieved from http://www.brunel.ac.uk/329/BBS%20documents/PHD%20Doctoral%20Symposium%2009/HassanAlMahdi7343.pdf

AL-Riyadh Newspaper. (2012). *The number of institutions and companies in Saudi Arabia more than the Saudis working in the private sector*. AL-Riyadh newspaper. Retrieved from http://riy.cc/725418

AMEinfo. (2008). *Saudi Arabia's retail sector is undergoing a major expansion*. Retrieved from http://www.ameinfo.com/ar-107430.html

Asfoura, E., Jamous, N., Kassam, G., & Dumke, R. (2009). E-Mall as Solution For Marketing The Federated ERP Components On The Basis of Web Services. *International Review of Business Research Papers, 5*(4), 478.

Auger, P., & Gallaugher, J. M. (1997). Factors affecting adoption of an Internet-based sales presence for small businesses. *The Information Society, 13*(1), 20.

Barry, H., & Milner, B. (2002). SME's and electronic commerce: A departure from the traditional prioritization of training? *Journal of European Industrial Training, 25*(7), 11.

Blili, S., & Raymond, L. (1993). Information Technology: Threats and Opportunities for Small and Medium-sized Enterprises. *International Journal of Information Management, 13*(6), 10. http://dx.doi.org/10.1016/0268-4012(93)90060-H

Bose, R., & Sugumaran, V. (2006). Challenges for Deploying Web Services-Based E-Business Systems in SMEs. *International Journal of E-Business Research, 2*(1), 18. http://dx.doi.org/10.4018/jebr.2006010101

Bundagji, F. Y. (2005). *Small Business and Market Growth in Saudi Arabia.* Retrieved from http://archive.arabnews.com/?page=6§ion=0&article=72204&d=24&m=10&y=2005

CITC (Communications and Information Technology Commission). (2010). *IT Report 2010 On the Internet Ecosystem in Saudi Arabia.* Communications and Information Technology Commission, Riyadh.

CSC. (2010). Retrieved from http://www.saudichambers.org.sa/2_9913_ENU_HTML.htm

Curbera, F., Khalaf, R., Mukhi, N., Tai, S., & Weerawarana, S. (2003). The next step in Web services. *Communications of the ACM, 46*(10), 6. http://dx.doi.org/10.1145/944217.944234

Deck, S. (1997). Ease of navigation key to successful Emalls. *Computerworld, 31*(28), 4, Retrieved from ABI/INFORM Global.

Dinesh, A. M., & Jaideep, M. (2001). Understanding small business electronic commerce adoption: An empirical analysis. *The Journal of Computer Information Systems, 41*(3), 70.

Dixon T., Thompson B., & McAllister P. (2002). *The value of ICT for SMEs in the UK: a critical literature review.* Report for Small Business Service research programme, The College of Estate Management.

Eastin, M. S. (2002). Diffusion of e-commerce: an analysis of the adoption of four e-commerce activities. *Telematics and Informatics, 19*(3), 251-267. http://dx.doi.org/10.1016/S0736-5853(01)00005-3

Fariselli, P., Oughton, C., Picory, C., & Sugden, R. (1999). Electronic commerce and the future for SMEs in a global market-place: Networking and public policies. *Small Business Economics, 12*(3), 15. http://dx.doi.org/10.1023/A:1008029924987

Fillis I., Johannson U., & Wagner B. (2004). Factors impacting on e-business adoption and development in the smaller firm. *International Journal of Entrepreneurial Behaviour & Research, 10*(2), 14.

Gable, G. (1991). Consultant Engagement for First Time Computerization: A Pro-active Client Role in Small Businesses. *Information & Management, 20*(2), 11.

Habtoor, A. (2011). *The retail sector: 600 thousand untapped jobs and 160 billion in the hands of foreign workers.* Retrieved from http://www.aleqt.com/2011/01/15/article_491563.html

Hamid, M. (2011). *Souq.com is dedicating e-commerce techniques.* Retrieved from http://www.arabianbusiness.com/arabic/602742?start=1

Koh, S., & Maguire, S. (2004). Identifying the adoption of e-business and knowledge management within SMEs. *Journal of Small Business and Enterprise Development, 11*(1), 11.

Laudon, K. C., & Traver, C. G. (2010). *E-commerce: business, technology, society* (6th ed.). Upper Saddle River, N.J: Prentice Hall

MacGregor R. C. (2004). Factors associated with formal networking in regional small business: some findings from a study of Swedish SMEs. *Journal of Small Business and Enterprise Development, 11*(1), 15. http://dx.doi.org/10.1108/14626000410519100

MCIT (Saudi Ministry of Communication and IT). (2011). *ICT indicators in K.S.A (H1-2011).* Retrieved from http://www.mcit.gov.sa/english/Development/SectorIndices/

Mintzberg, H. (1979). *The structuring of organization.* New jersey: Prantice Hall.

Mirchandani, D. A., & Motwani, J. (2001). Understanding small business electronic commerce adoption: An empirical analysis. *Journal of Computer Information Systems, 41*(3), 4.

Mustonen-Ollila, E. (1998). *Methodologies Choice and Adoption: Using Diffusion of Innovations as the Theoretical Framework.* Lappeenranta University of Technology, Lappeenranta.

Nooteboom, B. (1988). The Facts about Small Business and the Real Values of its life World. *American Journal of Economics and Sociolog, 47*(3), 16. http://dx.doi.org/10.1111/j.1536-7150.1988.tb02043.x

Qasrawi, N. (2001). *Environment and opportunities for e-commerce in the UAE* (2001 ed.). UAE: Albian Book.

Raosoft. (2004). *Sample size calculator.* Retrieved from http://www.raosoft.com/samplesize.html

Rogers, E. M. (1995). *Diffusion of innovations* (5th ed.). New York: Simon and Schuster.

Samiha F., & Sanjay L. (2002). *Globalization and firm competitiveness in the Middle East and North Africa Region* (illustrated ed.). World Bank Publications.

Sara C., & Dylan J. (2006). *Enterprise and Small Business: Principles, Practice and Policy* (2nd ed.). Financial Times Prentice Hall.

Scupola, A. (2002). *Adoption issues of business-to-business Internet commerce in European SMEs.* In Proceedings of the 35th Hawaii International Conference on System Sciences. Hawaii, USA. http://dx.doi.org/10.1109/HICSS.2002.994141

Susris. (2011). *A SME Authority for Saudi Arabia.* Retrieved from http://www.susris.com/2011/02/02/a-sme-authority-for-saudi-arabia/

Tan, Y. L., Macaulay, L. A., & Scheurer, M. (2006). Adoption of ICT among small business: vision vs. reality. *International Journal of Electronic Business, 5*(2), 16.

Taylor M., & Murphy A. (2004). SMEs and e-business. *Journal of Small Business and Enterprise Development, 11*(3), 10. http://dx.doi.org/10.1108/14626000410551546

Tetteh, E., & Burn, J. (2001). Global strategies for SME-business: Applying the SMALL framework. *Logistics Information Management, 14*(1-2), 10.

The Commission of the European Community. (2003). Concerning the definition of micro, small and medium-sized enterprises. *Official Journal of the European Union.* Retrieved from http://eur-lex.europa.eu/LexUriServ/LexUriServ.do?uri=OJ:L:2003:124:0036:0041:en:PDF

Thong, J. Y. L., Yap, C. W., & Raman, K. S. (1994). Engagement of External Expertise in Information Systems Implementation. *Journal of Management Information Systems, 11*(2), 23.

Timmers, P. (1998). Business Models for Electronic Markets. *Electronic Markets, 3*(2), 6.

Vézina M., Côté L., Sabourin V., & Pellerin M. (2003). *Electronic business models, a conceptual framework for small and medium-Sized Canadian enterprise.* CEFRIO.

Xu J., & Quaddus M. (2004). Adoption and diffusion of knowledge management systems: An Australian survey. *Journal of Management Development, 24*(4), 27.

Zimmermann, H. D. (1997). The model of regional electronic marketplaces--The example of the electronic Mall Bodensee (emb.net). *Telematics and Informatics, 14*(2), 117-130. http://dx.doi.org/10.1016/S0736-5853(96)00028-7